\documentclass[aps,groupedaddres8s,fleqn,nofootinbib,twocolumn]{revtex4}
\usepackage{graphicx}
\usepackage{xcolor}
\usepackage{amsmath,amssymb}
\usepackage{float}
\usepackage{amsfonts}
\usepackage{verbatim}
\usepackage{flushend}
\usepackage{balance}
\usepackage{endnotes}
\usepackage{footnote}
\usepackage{adjustbox}
\usepackage{gensymb}
\usepackage{float}
\usepackage{epigraph}
\usepackage{xcolor}
\usepackage[utf8]{inputenc}
\usepackage{setspace}

\newcommand{\beq}{\begin{eqnarray}}
\newcommand{\eeq}{\end{eqnarray}}
\usepackage{amsmath}

\begin{document}

\author{K. Trachenko$^1$}
\address{$^1$ School of Physical and Chemical Sciences, Queen Mary University of London, Mile End Road, London, E1 4NS, UK}

\title{Theory of melting lines}

\begin{abstract}
Our understanding of the three basic states of matter (solids, liquids and gases) is based on temperature and pressure phase diagrams with three phase transition lines: solid-gas, liquid-gas and solid-liquid lines. There are analytical expressions $P(T)$ for the first two lines derived on a purely general-theoretical thermodynamic basis. In contrast, there exists no similar function for the third, melting, line (ML). Here, we develop a general two-phase theory of MLs and their analytical form. This theory predicts the parabolic form of the MLs for normal melting, relates the MLs to thermal and elastic properties of liquid and solid phases and quantitatively agrees with experimental MLs in different system types. We show that the parameters of the ML parabola are governed by fundamental physical constants. In this sense, parabolic MLs possess universality across different systems.
\end{abstract}

\maketitle

\section{Introduction}

In statistical physics and thermodynamics, coexistence of different phases is set by the equality of thermodynamic potentials. This results in the Clausius-Clapeyron (CC) equation describing phase transition lines on the phase diagram:

\begin{equation}
\frac{dP}{dT}=\frac{S_1-S_2}{V_1-V_2}
\label{e1}
\end{equation}

\noindent where $\frac{dP}{dT}$ is the slope of the transition line and $S$ and $V$ are entropy and volume of any two phases \cite{landaustat,kittel,chemreview}.

Eq. \eqref{e1} readily gives important results for two transition lines, the sublimation and boiling lines \cite{landaustat,kittel}. This involves two steps. First, the gas volume is much larger than volume of solids and liquids so that $V_1-V_2$ becomes the gas volume obeying the gas equation of state (EoS). Second, assuming the constant latent heat $q=T(s_1-s_2)$ in Eq. \eqref{e1} ($s_i$ are entropies per particle) and integrating gives a purely theoretical line of saturated vapor pressure along the sublimation and boiling lines as

\begin{equation}
P(T)\propto e^{-\frac{q}{T}}
\label{plines}
\end{equation}

Eq. \eqref{plines} is a textbook function for the sublimation and boiling lines \cite{landaustat,kittel}. It is broadly considered to be in agreement with experiments \cite{kittel,walasexp}, although cases of disagreement are discussed too \cite{reidexp}.

In contrast, there is no analytical function $P(T)$ derived for the third, melting, line of a theoretical generality comparable to sublimation and boiling lines. This generality includes not relying, as was done in the past, on phenomenological fitting functions, arbitrary series truncations, assumptions of a particular system-specific melting mechanism, using computer modelling to calculate the melting line (ML), its parts or related data, choosing specific liquid structure or interactions and so on \cite{chemreview,ubbelohde,anderson1,proctor2}.

We note that the ML is not limited by triple and critical points from above as the other two transition lines are. Continuing up to very high temperature where system properties change due to, e.g., ionisation, the ML extends over a much larger range of the phase diagram.

Understanding {\it mechanisms} of melting has a long history \cite{ubbelohde,anderson1,mybook,chemreview}. Starting from 1913, Sommerfeld and Brillouin aimed to develop a thermodynamic theory of liquids based on phonons \cite{brillouin,somm,br1}. This was around the time when phonon-based theories of Einstein and Debye were published \cite{einstein,debyepaper} laying the foundations of the solid state theory. Sommerfeld and Brillouin considered that, differently from solids, liquids do not support shear stress and shear phonons \cite{brillouin,somm,br1}. Brillouin \cite{brilprb} and independently Born \cite{bornmelting} then defined melting as the point where solid's rigidity disappears. This elicited an objection from Frenkel \cite{frenkel}: liquids too support shear stress, albeit at high frequency. This is a valid point to which we return to later. Other mechanisms include semi-phenomenological Lindemann-Gilvarry criteria \cite{lindemann,gilvarry} based on threshold atomic displacements a solid can tolerate before melting. These criteria continue to remain of interest \cite{khrapak}, although they are known to agree with experiments in some systems only but not others \cite{ubbelohde,anderson1}. Other approaches include defect-mediated mechanisms with system-specific defects such as dislocations, surfaces and premelting \cite{ubbelohde,anderson1,chemreview}. It is therefore unlikely that a microscopic theory based on a melting mechanism exists that is generally applicable. We also note that the earlier criteria above are one-phase approaches and focus on the solid phase only, whereas both solid and liquid phases are equally relevant for understanding melting which is set by the equality of thermodynamic potentials of the two phases as reflected in the CC equation \eqref{e1} \cite{landaustat,ubbelohde,chemreview}.

Faced with these historical issues, we can nevertheless aim for a general thermodynamic theory of the MLs which is independent of a particular melting mechanism and where system-specific effects contribute to thermodynamic potentials. This is of importance for general theory as well as applications: predicting phase diagrams is an active research extending to materials science, with the need to understand liquids emphasised \cite{calphad1,calphad2}. As discussed in the next section, developing a general theory of MLs faced, until recently, fundamental problems of liquid description.

Here, we develop a general two-phase theory of MLs and their analytical form. This theory predicts the parabolic form of the MLs for normal melting, relates the MLs to thermal and elastic properties of solid and liquid phases and agrees with experiments in different system types including noble, molecular, network and metallic. We show that the parameters of the ML parabola are governed by fundamental physical constants. In this sense, parabolic melting lines possess universality across different systems.

\section{Results and Discussion}

\subsection{The main idea: liquid energy}

The main reason for the absence of a general analytical function for the ML is related to fundamental problems involved in liquid theory. In this section, we recall the origin of these problems. This is followed by the summary of how these problems are overcome in the phonon theory of liquid thermodynamics. This theory gives an equation for the liquid energy which will be used in the next section to derive the differential equation for the ML.

Deriving the ML theoretically is faced with several fundamental problems, explaining why it was not achieved before. First, $q$ for the sublimation and boiling lines is related to transitions between a condensed solid or liquid state and the uncondensed gas state. For this reason, $q$ includes the cohesive energy of the condensed state and can be considered roughly constant. This does not apply to the ML because it separates the states which are both condensed. Hence one needs to explicitly know the liquid entropy in order to solve the CC equation \eqref{e1} for the ML. Differently from solids and gases, liquid thermodynamic properties including entropy are not generally known from theory \cite{landaustat}. Second, one needs to explicitly know the liquid $V$ as a function of $P$ and $T$ and hence the liquid EoS in order to solve the CC equation \eqref{e1}. This EoS is not generally known from theory \cite{landaustat}, differently from the gas EoS used to derive the boiling and sublimation lines from Eq. \eqref{e1}. Finally, $V_1-V_2$ in Eq. \eqref{e1} can't be approximated by $V$ of either phase as is done when deriving the sublimation and boiling lines because solids and liquids are both condensed states with relative differences of $V$ of about 10\% only \cite{ubbelohde}.

These problems, the unavailability of a general theory of liquid thermodynamic properties including entropy and EoS, are part of a more general fundamental problem of liquid theory. As discussed by Landau and Lifshitz (LL) \cite{landaustat}, the simplifying features of solids and gases are small oscillations and weak interactions, respectively. Liquids have neither because atomic displacements are large due to flow and interactions are as strong as in solids. This fundamental problem is summarised to say that liquid theory has no small parameter \cite{pitaevskii}. Because the interactions are strong, all thermodynamic properties strongly depend on the type of interactions and therefore are strongly system-dependent. This, conclude LL \cite{landaustat}, rules out a generally applicable theory of liquid thermodynamic properties and their temperature dependence (this assertion was consistent with Landau's own view according to Peierls \cite{peierls-frenkel}). These fundamental problems extend to all theories based on explicitly considering liquid interactions and structure, the mainstream approach to liquid physics in the last century \cite{borngreen,green2,kirkwoodbook,fisher,barkerhenderson,egelstaff,hansen1,hansen2,march,faber,wca1,wca2,rosentar,zwanzig}. This includes expanding interactions into short-range repulsive reference and attractive terms in simple models (see, e.g., \cite{wca1,wca2,rosentar,zwanzig}): the expansions and their coefficients are system-dependent and the results are not generally applicable. Apart from research, these problems impacted teaching \cite{granato}.
Importantly, these issues do not emerge in the solid state theory: this theory operates in terms of excitations in the system, phonons, rather than considers interactions and structure explicitly \cite{mybook}. As a result, the derived thermodynamic properties are applicable to all solids \cite{landaustat}.

The liquid entropy in the CC equation \eqref{e1} is particularly hard to approach head on: it includes vibrational and configurational terms, and its unclear how to balance the two \cite{ubbelohde,dyre,ropp}. This entropy is not available, apart from simple models such as hard-sphere or van der Waals systems used to discuss liquids \cite{parisihard,parisi}. These models give the specific heat $c_v=\frac{3}{2}k_{\rm B}$ \cite{landaustat,wallacecv,wallacebook,mybook} and therefore describe the ideal gas thermodynamically. This is far removed from real liquids where $c_v\approx 3k_{\rm B}$ close to the ML \cite{wallacecv,wallacebook,ropp,mybook,nist} which we are interested in and where $c_v$ is as large as in solids.

Our main idea to overcome the problem of solving the CC equation \eqref{e1} for the ML is to write this equation in terms of energy instead of entropy. The liquid energy is known on the basis of the recent phonon theory of liquids. This theory \cite{ropp,mybook,proctor1,proctor2,chen-review} is based on the main premise of statistical physics: properties of an interacting system are governed by its excitations \cite{landaustat,landaustat1}. In solids, these are collective excitations, phonons. The theory further focuses on energy $E$ as the primary property in statistical physics \cite{landaustat}, and this makes the liquid problem tractable (entropy can be found by integrating $\frac{1}{T}\frac{dE}{dT}$ if needed). In view of long-lasting fundamental problems of liquid theory above, it is useful to briefly summarise the theory and its main result which we will use.

The key to addressing the fundamental small parameter problem stated by Landau, Lifshitz and Pitaevskii is that this parameter does exist in liquids and is the same as solids, small phonon displacements (this is grounded in microscopic liquid dynamics: between diffusive jumps enabling liquid flow, liquid particles vibrate around quasi-equilibrium positions \cite{frenkel,dyre,ropp}). In an important difference to solids where the phase space available to phonons is fixed, this phase space in liquids is {\it variable}  \cite{ropp,mybook}.

Experiments have shown that propagating phonons in liquids and solids are remarkably similar in terms of dispersion curves \cite{copley,pilgrim,burkel,pilgrim2,water-tran,hoso,hoso3,monaco1,monaco2,sn,ropp,mybook}. The phonon theory of liquid thermodynamics \cite{ropp,mybook,proctor1,proctor2,withbook} further observes that transverse phonons in liquids exist above the gap in the reciprocal space $k_g=\frac{1}{c\tau}$ \cite{yangprl,ropp,gmsreview,hosokawa-gap} only, where $\tau$ is the liquid relaxation time identified with the average hopping time of molecules from one quasi-equilibrium point to the next and $c$ is the speed of sound \cite{frenkel,dyre,mybook}. This is equivalent to propagating phonons (in a sense of frequency exceeding the decay rate) above the hopping frequency $\omega_{\rm F}=\frac{1}{\tau}$ \cite{ropp1,mybook} as envisaged by Frenkel originally \cite{frenkel}. This gives the variability of the phonon phase space in liquids. There are several ways to measure the hopping frequency $\omega_{\rm F}$, including using the approximate relation $\omega_{\rm F}=\frac{G}{\eta}$, where $\eta$ is viscosity and $G$ is the high-frequency shear modulus \cite{frenkel,dyre}.

The energy of phonons excitations in the liquid is the sum of the longitudinal phonon energy and the energy of propagating gapped transverse phonons \cite{ropp,ropp1,mybook}. Differently from solids, liquids have another type of excitations: these are diffusing atoms which enable liquid flow. Adding the energy of this motion to the phonon energy gives the liquid energy as \cite{ropp,mybook,proctor1,proctor2,withbook}

\begin{equation}
E_l=NT\left(3-\left(\frac{\omega_{\rm F}}{\omega_{\rm D}}\right)^3\right)
\label{harmo}
\end{equation}

\noindent where $\omega_{\rm D}$ is the maximal Debye frequency that sets the high-temperature limit for $\omega_{\rm F}$. Here and below, $k_{\rm B}=1$.

Eq. \eqref{harmo} predicts that liquid specific heat $c_v=\frac{1}{N}\frac{dE}{dT}$ is

\begin{equation}
c_v\approx 3
\end{equation}

\noindent at the melting point where $\frac{\omega_{\rm F}}{\omega_{\rm D}}$ is small (see next section for a detailed discussion). Therefore, Eq. \eqref{harmo} predicts that liquid $c_v$ is close to that in solids in this regime.

$c_v\approx 3$ in liquids close to melting is in agreement with experiments in many liquids with different structure and bonding types \cite{wallacecv,wallacebook,ropp,mybook,proctor1,proctor2,nist}. Fig. 1 of Ref. \cite{wallacecv} and Fig. 9 of Ref. \cite{ropp} quickly make this point.

As $\omega_{\rm F}$ increases with temperature, $E_l$ changes from $3NT$ to $2NT$ when $\omega_{\rm F}$ reaches $\omega_{\rm D}$ at the Frenkel line where transverse phonons disappear from the liquid spectrum \cite{flreview} (recall that the absence of transverse phonons was seen as a general property of liquids by Sommerfeld and Brillouin \cite{brillouin,somm,br1}). This change of $E_l$ gives the {\it decrease} of $c_v$ from $3$ to $2$. This decrease is universally seen in liquids and is related to the reduction of the phonon phase space \cite{ropp,proctor1,proctor2,mybook,nist}.

Eq. \eqref{harmo} and its extensions to quantum effects \cite{ropp,mybook} have undergone independent ``detailed and rigorous'' tests against experiments for different types of liquids in a wide range of $T$ and $P$ \cite{proctor1,proctor2}. Noting that the theory is falsifiable and pursuing this falsifiability, these tests concluded that the theory is supported by experiments in a number of ways.

\subsection{Differential equation for the melting line}

We write the CC equation \eqref{e1} as

\begin{equation}
(V_l-V_s)\frac{dP}{dT}=(S_l-S_s)
\label{e101}
\end{equation}

\noindent where indexes $l$ and $s$ refer to liquids and solids from now on, differentiate \eqref{e101} and multiply the result by $T$:

\begin{equation}
T\frac{d^2P}{dT^2}(V_l-V_s)+T\frac{dP}{dT}\left(\frac{dV_l}{dT}-\frac{dV_s}{dT}\right)=T\frac{dS_l}{dT}-T\frac{dS_s}{dT}
\label{e2}
\end{equation}

The derivatives $\frac{dV_{l,s}}{dT}$ in Eq. \eqref{e2} are along the ML. Below we will relate the ML and its properties to thermal and elastic characteristics of the solid and liquid phases. This includes the thermal expansion coefficient $\alpha$ and the bulk modulus $B$ calculated from separate experimental data. This data enables us to calculate $\frac{dV}{dT}$ in Eq. \eqref{e2} along the ML directly. However, it is useful to relate $\frac{dV}{dT}$ to the standard definition of $\alpha=\frac{1}{V}\left(\frac{\partial V}{\partial T}\right)_P$ taken at constant pressure: $\frac{dV}{dT}=\left(\frac{\partial V}{\partial T}\right)_P+\left(\frac{\partial V}{\partial P}\right)_T\frac{dP}{dT}=\alpha V-\frac{V}{B}\frac{dP}{dT}$, where $B$ is the isothermal bulk modulus. Similarly, $\frac{dS}{dT}$ in Eq. \eqref{e2} is $\frac{dS}{dT}=
\left(\frac{\partial S}{\partial T}\right)_P+\left(\frac{\partial S}{\partial P}\right)_T\frac{dP}{dT}=\left(\frac{\partial S}{\partial T}\right)_P-\left(\frac{\partial V}{\partial T}\right)_P\frac{dP}{dT}=\left(\frac{\partial S}{\partial T}\right)_P-\alpha V\frac{dP}{dT}$. Using these $\frac{dV}{dT}$ and $\frac{dS}{dT}$ in Eq. \eqref{e2} gives

\begin{eqnarray}
\begin{split}
& T\frac{d^2P}{dT^2}(V_l-V_s)+2T\frac{dP}{dT}(\alpha_lV_l-\alpha_sV_s)-\\
& T\left(\frac{V_l}{B_l}-\frac{V_s}{B_s}\right)\left(\frac{dP}{dT}\right)^2=T\left(\frac{\partial S_l}{\partial T}\right)_P-T\left(\frac{\partial S_s}{\partial T}\right)_P
\label{e3}
\end{split}
\end{eqnarray}

The right-hand side is the difference between the constant-pressure heat capacities $C_p$ of solids and liquids across the ML. $C_p$ is related to the constant-volume heat capacity $C_v=\frac{\partial E}{\partial T}$, where $E$ is energy and the derivative is at constant volume, as \cite{landaustat}

\begin{equation}
T\left(\frac{\partial S}{\partial T}\right)_P=\frac{\partial E}{\partial T}+VT\alpha^2 B
\label{cpcv}
\end{equation}

Using this in Eq. \eqref{e3} gives the equation depending on energies $E_l$ and $E_s$ of both phases as

\begin{widetext}
\begin{equation}
T\frac{d^2P}{dT^2}(V_l-V_s)+2T\frac{dP}{dT}(\alpha_lV_l-\alpha_sV_s)-T\left(\frac{V_l}{B_l}-\frac{V_s}{B_s}\right)\left(\frac{dP}{dT}\right)^2=\frac{\partial}{\partial T}\left(E_l-E_s\right)+T(V_l\alpha_l^2B_l-V_s\alpha_s^2B_s)
\label{e31}
\end{equation}
\end{widetext}

We now recall the energies $E_s$ and $E_l$: $E_s=3NT$ and $E_l$ in Eq. \eqref{harmo}. Using $E_s$ and $E_l$ in Eq. \eqref{e31} gives

\begin{widetext}
\begin{equation}
T\frac{d^2P}{dT^2}(V_l-V_s)+2T\frac{dP}{dT}(\alpha_lV_l-\alpha_sV_s)-T\left(\frac{V_l}{B_l}-\frac{V_s}{B_s}\right)\left(\frac{dP}{dT}\right)^2=-N\left(\frac{\omega_{\rm F}}{\omega_{\rm D}}\right)^3\left(1+\frac{3\partial\ln\omega_{\rm F}}{\partial\ln T}\right)+T(V_l\alpha_l^2B_l-V_s\alpha_s^2B_s)
\label{e4}
\end{equation}
\end{widetext}

All terms in Eq. \eqref{e4} scale with $N$ and $V$. However, the first term on the right-hand side is much smaller than all other terms and can be dropped because $\frac{\omega_{\rm F}}{\omega_{\rm D}}\ll 1$ at the ML. Recalling $\omega_{\rm F}=\frac{G}{\eta}$ gives $\frac{\omega_{\rm F}}{\omega_{\rm D}}$ at the ML as

\begin{equation}
\frac{\omega_{\rm F}}{\omega_{\rm D}}\approx\frac{\eta_0}{\eta(T=T_m)}
\label{ratio}
\end{equation}

\noindent where $T_m$ is the melting temperature and $\eta_0$ is the high-temperature limiting value of $\eta$.

In viscous melts, the ratio \eqref{ratio} is very small. In a common system SiO$_2$, $\eta(T=T_m)$ is about 10$^6$ Pa$\cdot$ s \cite{ojovan}. Typical values of $\eta_0$ are 10$^{-5}$-10$^{-4}$ Pa$\cdot$ s \cite{nussinov,sciadv} and are close to $\eta\approx 10^{-3}$ Pa$\cdot$s in simulated SiO$_2$ where $\eta$ saturates to its high-temperature constant \cite{horbach}. This gives $\frac{\omega_{\rm F}}{\omega_{\rm D}}$ in Eq. \eqref{ratio} of about $10^{-9}$ and similarly small values in other viscous melts.

$\frac{\omega_{\rm F}}{\omega_{\rm D}}\ll 1$ also applies to low-viscous liquids such as water. Water viscosity at ambient conditions is interestingly close to the minimal quantum viscosity \cite{pt2021,myreview}, the lowest viscosity that a liquid can ever attain \cite{sciadv}. Yet even in this case, $\frac{\omega_{\rm F}}{\omega_{\rm D}}$ is small. Experimental $\frac{\eta_0}{\eta(T=T_m)}\approx 0.16$ in room-pressure water \cite{nist}. This is consistent with X-ray scattering experiments showing the viscoelastic behavior of water in a wide temperature range ($T_m, T_m+100$ K) where molecules oscillate many times before jumping to new quasi-equilibrium positions \cite{water}, implying $\omega_{\rm F}\ll{\omega_{\rm D}}$. At 10, 100 and 1000 MPa, experimental $\frac{\eta_0}{\eta(T=T_m)}=\frac{\omega_{\rm F}}{\omega_{\rm D}}$ are in the range 0.03-0.06 \cite{nist}. These small ratios $\frac{\omega_{\rm F}}{\omega_{\rm D}}$ enter as a cube in Eq. \eqref{e4}.

The smallness of $\frac{\omega_{\rm F}}{\omega_{\rm D}}$ in liquids at the ML is in perfect agreement with the experimental $c_v$ of many liquids. According to Eq. \eqref{harmo}, $\frac{\omega_{\rm F}}{\omega_{\rm D}}\ll 1 $ gives the liquid $c_v$ close to its solid value, $c_v=3$. This $c_v$ is seen in liquids close to the ML in a wide range of temperature and pressure and includes liquids with different structure and bonding types: noble, metallic, semiconducting and molecular \cite{wallacecv,wallacebook,ropp,mybook,proctor1,proctor2,nist}. This experimental fact supports the smallness of the energy difference term $E_l-E_s$ in Eq. \eqref{e31} and of $\frac{\omega_{\rm F}}{\omega_{\rm D}}$ in Eq. \eqref{e4}.

Physically, this smallness of the energy difference $E_l-E_s$ at the ML is due to solids and liquids being both condensed states of matter where the characteristic scales of energy and interatomic separation are set by the Rydberg energy and Bohr radius involving fundamental physical constants \cite{myreview}. The closeness of most important system properties then follows \cite{myreview}. This does not apply to phases separated by the other two transition lines, sublimation and boiling lines, because they involve the non-condensed gas phase where the Rydberg energy and Bohr radius do not operate.

This last point makes an interesting connection to our discussion in the previous section where we noted one of the problems of solving the CC equation \eqref{e1} for the ML: it can not be assumed that the latent heat $q$ includes a large contribution from the cohesive energy of solids (liquids) as is the case for sublimation (boiling) lines. Indeed, solids and liquids are both condensed phases, and $q$ is related to the system-specific difference of cohesive energies of the two states. In our theory here, the condensed nature of solids and liquids is turned into an advantage: the closeness of energies of these states in Eq. \eqref{e31}, $E_l$ and $E_s$, the consequence of both states being condensed, simplifies the theory.

For $\frac{\omega_{\rm F}}{\omega_{\rm D}}\ll 1$, Eq. \eqref{e4} reads

\begin{eqnarray}
\begin{split}
& \frac{d^2P}{dT^2}(V_l-V_s)+2\frac{dP}{dT}(\alpha_lV_l-\alpha_sV_s)-\\
&\left(\frac{V_l}{B_l}-\frac{V_s}{B_s}\right)\left(\frac{dP}{dT}\right)^2=V_l\alpha_l^2B_l-V_s\alpha_s^2B_s
\label{e10}
\end{split}
\end{eqnarray}

This second-order equation does not suggest a link to a second-order transition \cite{landaustat}. We are considering the first-order melting transition with parameters which enter in Eq. \eqref{e10} through the entropy derivative in Eq. \eqref{cpcv}.

As set out in the previous section, Eq. \eqref{e10} eliminates the problem of knowing the liquid entropy and depends on thermal and elastic properties of solids and liquids only. At the same time, Eq. \eqref{e10} is quite general and follows directly from the CC equation \eqref{e1} using one assumption only: the equality of $c_v$ (or energy) of solids and liquids across the ML. As discussed earlier, this equality is backed up by theory and numerous experiments.

Eq. \eqref{e10} predicts the function $P(T)$ of the ML. The parameters of this function are set by the change of thermal and elastic properties of solids and liquids on crossing the ML. This naturally makes our theory a {\it two-phase} theory and different to earlier one-phase theories which considered how the properties of solids change on melting and left liquids out from consideration. These theories include those of Born, Brillouin, Lindemann-Gilvarry and so on \cite{ubbelohde,anderson1,chemreview,proctor2,bornmelting,frenkel,brilprb,lindemann,gilvarry}. These one-phase approaches focusing solely on solids came about because solids were understood whereas liquids weren't as discussed in the previous section. However, there is no physical reason to single out solids to understand the transition between solids and liquids: both phases are equally relevant to understand this transition which is set by the equality of thermodynamic potentials of the two phases \cite{landaustat,ubbelohde}. Recall that the two-phase approach is implicit in deriving the sublimation and boiling lines where the latent heat $q=T\Delta s$ reflects the entropy change between two phases and enters the analytical form \eqref{plines} as a parameter.

Eq. \eqref{e10} provides a relation between different properties: thermal and elastic properties of solids and liquids on one hand and melting lines on the other hand. Providing such relations is viewed as an essence of a physical theory \cite{landaupeierls}.

We note in that Eq. \eqref{harmo} assumes that the energy of each contributing phonon is given by $T$ in the harmonic approximation. Accounting for phonon anharmonicity modifies $E_l$ and $E_s$ as \cite{ropp,mybook}:

\begin{eqnarray}
\begin{split}
& E_l^a=NT\left(1+\frac{\alpha_lT}{2}\right)\left(3-\left(\frac{\omega_{\rm F}}{\omega_{\rm D}}\right)^3\right)\\
& E_s^a=3NT\left(1+\frac{\alpha_sT}{2}\right)
\label{anharmo}
\end{split}
\end{eqnarray}

\noindent where setting the hopping frequency $\omega_{\rm F}=0$ in the liquid corresponds to the solid and $E_l^a=E_l^s$ as expected.

Using $E_l^a-E_s^a$ in the energy difference in Eq. \eqref{e31} and $\frac{\omega_{\rm F}}{\omega_{\rm D}}\ll 1$ as before contributes the extra term $3N(\alpha_l-\alpha_s)$ to the right-hand side of Eq. \eqref{e10}. Using the Gr\"{u}neisen parameter $\gamma=\frac{V\alpha B}{C_v}\approx\frac{V\alpha B}{3N}$ \cite{anderson1}, this term is $\frac{V_l\alpha_l^2B_l}{\gamma_l}-\frac{V_s\alpha_s^2B_s}{\gamma_s}$ or about $V_l\alpha_l^2B_l-V_s\alpha_s^2B_s$ in view of $\gamma\approx 1$. This approximately doubles the term on the right-hand side of Eq. \eqref{e10}.

\subsection{Solution, its properties and the parabolic form of the melting line}
\label{solution}

Written in terms of functions $a$, $b$ and $d$, Eq. \eqref{e10} is

\begin{eqnarray}
\begin{split}
& \frac{d^2P}{dT^2}+a\frac{dP}{dT}-b\left(\frac{dP}{dT}\right)^2=d\\
& a=\frac{2(\alpha_lr-\alpha_s)}{r-1}\\
& b=\left(\frac{r}{B_l}-\frac{1}{B_s}\right)\frac{1}{r-1}\\
& d=\frac{\alpha_l^2B_lr-\alpha_s^2B_s}{r-1}\\
& r=\frac{V_l}{V_s}
\label{e11}
\end{split}
\end{eqnarray}

\noindent where $r$ is the ratio of the liquid and solid volumes and is typically about 1.1 \cite{ubbelohde} (the relative volume increase at melting is typically 10-15\% for noble and molecular systems, 1-5\% in monoatomic metals and semiconductors including negative values and 10-20\% for binary salts \cite{ubbelohde}).

In general, $a$, $b$ and $d$ depend on $P$ and $T$ because $\alpha$, $B$ and $r$ do. In this case, the ML comes from solving Eq. \eqref{e11} numerically. $a(P,T)$, $b(P,T)$ and $d(P,T)$ can be taken from (a) experiments or modelling data including the EoS or (b) combining this data at low pressure with models predicting the variation of $\alpha$ and $B$ with $P$ and $T$ \cite{anderson1}. In the latter case, analytical solutions of Eq. \eqref{e11} might be possible depending on the model.

To analytically solve Eq. \eqref{e11} in order to check theory predictions, we make two approximations. These approximations are independent and sequentially interchangeable. First, we assume that $\alpha$, $B$ and $r$ do not substantially change for sufficiently small variations of $P$ and $T$ and can be approximately considered constant \cite{anderson1}. This assumption is analogous to considering the constant latent heat $q$ in the derivation of the boiling and sublimation lines from the CC equation \cite{landaustat,kittel}. In liquids, $\alpha$ and $B$ depend on both viscous and elastic components of motion \cite{frenkel,prbtg}. The viscous component is governed by the activation energy that needs to be surmounted by the jumping molecule, $U$. At small pressure, $U=8\pi Gr\Delta r^2$, where $G$ is the liquid high-frequency shear modulus, $r$ is the cage radius and $\Delta r$ is the temperature-induced fluctuating increase of the cage size that enables the molecule to escape the cage \cite{frenkel,dyre}. Pressure typically increases $U$ (unless there is an anomalous structural change from, for example, covalent to molecular liquid). This increase is the extra work needed to expand the cage from radius $r$ to $r+\Delta r$, $4\pi r^2\Delta r P$. Then, $U$ becomes

\begin{equation}
U=4\pi r\Delta r(G\Delta r+Pr)
\label{activ}
\end{equation}

The internal elastic and external pressure effects become comparable when $P=G\frac{\Delta r}{r}$. Taking $\Delta r\lesssim r$ gives $P$ on the order of GPa in noble and molecular systems and 10 GPa in stiffer systems such as Fe. A similar estimation of the variation of $B$ due to the elastic component can be done by writing $B=B_0+\frac{\partial B}{\partial P}P$, where $B_0$ is the bulk modulus at zero pressure and $\frac{\partial B}{\partial P}$ is typically 4-6 for different minerals and oxides \cite{anderson1}. In these systems, high-temperature $\alpha$ typically increases by about 3\% over 100 K at ambient pressure.

We add two remarks regarding variation of $\alpha$, $B$ and $r$ in Eq. \eqref{e11}. First, pressure and temperature have competing effects on $\alpha$ and $B$ (pressure usually increases $B$ and decreases $\alpha$, whereas temperature does the opposite). This gives slower variation of $\alpha$ and $B$ along the ML if pressure along the line increases with temperature as it usually does. Second, Eq. \eqref{e11} contains $V_l$ and $V_s$ in addition to $\alpha$ and $B$, however $V_l$ and $V_s$ enter Eq. \eqref{e11} as the ratio $r=\frac{V_l}{V_s}$ which varies slower than the volumes themselves. For example, this ratio is constant along the ML for the inverse power potentials \cite{stishovmelt} (see Section \ref{otherforms}).

Considering constant $\alpha$, $B$ and $r$ gives constant $a$, $b$ and $d$ in Eq. \eqref{e11}. The solution of Eq. \eqref{e11} is then

\begin{eqnarray}
\begin{split}
& P=P_0+\frac{a}{2b}T-\frac{1}{b}\ln\cos\left(\sqrt{\Delta}T\right) \\
& \sqrt{\Delta}=\sqrt{bd-\frac{a^2}{4}}
\label{sol}
\end{split}
\end{eqnarray}

\noindent where $P_0$ is the integration constant which can be used to fix the location of the triple point for a particular system and we set the second integration constant in the argument of $\cos$ to zero.

We consider the most general case taking phonon anharmonicity into account. As discussed in the previous section, this anharmonicity doubles $d$ in Eq. \eqref{e11}. Using $a$, $b$ and $2d$ in Eq. \eqref{e11} gives $\Delta=2bd-\frac{a^2}{4}$ as

\begin{equation}
\Delta=\frac{1}{\left(r-1\right)^2}\left((\alpha_lr-\alpha_s)^2-2r\left(\alpha_l\sqrt{\frac{B_l}{B_s}}-\alpha_s\sqrt{\frac{B_s}{B_l}}\right)^2\right)
\label{delta}
\end{equation}

In common non-anomalous cases where $\alpha_l>\alpha_s$, $B_s>B_l$ (the reason these inequalities often apply is discussed later) and with typical $r$, $\sqrt{\Delta}$ is real. This is confirmed by using the actual values of $\alpha$ and $B$ for liquids and solids \cite{prbtg} and a range of $r$ in Eq. \eqref{delta}.

%We note that not accounting for phonon anharmonicity and considering a purely harmonic case with $a$, $b$ and $c$ in Eq. \eqref{e11} gives negative $\Delta$:

%\begin{equation}
%\Delta=-\frac{r}{\left(r-1\right)^2}\left(\alpha_l\sqrt{\frac{B_l}{B_s}}-\alpha_s\sqrt{\frac{B_s}{B_l}}\right)^2
%\end{equation}
%\noindent and the $\cos$ function in Eq. \eqref{sol} transforms into $\cosh$.

Eq. \eqref{sol} makes several predictions regarding the ML. First, below we will consider normal cases where $\alpha_l>\alpha_s$, $B_s>B_l$ and $r>1$. In this case, the coefficient in front of the linear term in Eq. \eqref{sol}, $\frac{a}{2b}$, is positive according to Eq. \eqref{e11}. Second, the last term in Eq. \eqref{sol} with real $\sqrt{\Delta}$ in normal systems gives the increase of the slope of the $P(T)$ melting line, $\frac{dP}{dT}$, with temperature. We will discuss experimental data showing this increase below. However, we note that a combination of anomalous $r$, $\alpha$ and $B$ in Eq. \eqref{delta} can give imaginary $\sqrt{\Delta}$. In this case, $\cos\left(\sqrt{\Delta}T\right)$ in Eq. \eqref{sol} becomes $\cosh\left(\sqrt{|\Delta|}T\right)$, resulting in the decrease of $\frac{dP}{dT}$ with temperature as seen in anomalous cases such as Ce \cite{proctor2}. $\frac{dP}{dT}$ in Eq. \eqref{sol} can also become negative (with $\frac{a}{2b}$ being either positive or negative if $\alpha_l<0$ in Eq. \eqref{e11}) as seen in anomalous systems such as water or in systems with re-entrant melting \cite{chemreview,reentrant}. $\frac{dP}{dT}$ can also change its sign, corresponding to a non-monotonic behavior of the ML \cite{chemreview}. These different scenarios of anomalous melting are described by the theory, however the detailed discussion of these scenarios is outside the scope of this paper. Here, we discuss commonly encountered normal melting where $\frac{dP}{dT}$ is positive and increases with temperature.

Third, $\sqrt{\Delta}$ in Eq. \eqref{delta} is on the order of $\alpha_l$ (calculating $\sqrt{\Delta}$ using the real values of $\alpha$ and $B$ in liquids and solids \cite{prbtg} and $r=1.1$ gives $\sqrt{\Delta}\approx 4\cdot(10^{-4}-10^{-3})$K$^{-1}\approx (4-9)\alpha_l$ for different systems). Hence $\sqrt{\Delta}T$ is small in the temperature range of MLs considered later in Fig. 1. For small $\sqrt{\Delta}T$, Eq. \eqref{sol} reads

\begin{equation}
P=P_0+\frac{a}{2b}T+\frac{\Delta}{2b}T^2
\label{sol1}
\end{equation}

In the temperature range set by the approximation above, Eq. \eqref{sol1} predicts that the melting line is a {\it parabola}. This is an attractive function in view of its simplicity and common occurrence in physics. For a given pressure, Eq. \eqref{sol1} predicts the temperature of melting.

Fourth, the second term in Eqs. \eqref{sol} and \eqref{sol1} predicts the value of the linear slope contributing to the overall $\frac{dP}{dT}$ of the ML and the last term predicts the deviation from linearity. Further insight into these values follows from making the second approximation to $a$, $b$ and $d$ in Eq. \eqref{e11}. $\alpha_l$ is often substantially larger than $\alpha_s$ and $B_l$ is substantially smaller than $B_s$. This is because the temperature and elastic response includes both elastic and viscous components in liquids and only an elastic component in solids \cite{frenkel,prbtg} (for example, typical ranges of $\alpha_l$ and $\alpha_s$ are $(10^{-5}-10^{-3}$) K$^{-1}$ and ($10^{-6}-10^{-5}$) K$^{-1}$, respectively). This gives $a=\frac{2\alpha_lr}{r-1}$, $b=\frac{r}{B_l}\frac{1}{r-1}$ in Eq. \eqref{e11}, $\frac{a}{2b}=\alpha_lB_l$ and $\frac{\Delta}{2b}=\frac{r}{2(r-1)}\alpha_l^2B_l$. We can check these approximations to $\frac{a}{2b}$ and $\frac{\Delta}{2b}$ by using $\alpha$ and $B$ in liquids and solids \cite{prbtg} and the typical value $r=1.1$ \cite{ubbelohde}. We find that $\alpha_lB_l$ underestimates $\frac{a}{2b}$ by a factor of 1.4 on average and $\frac{r}{2(r-1)}\alpha_l^2B_l$ overestimates $\frac{\Delta}{2b}$ by 1.7 on average while the order of magnitude of $\frac{a}{2b}$ and $\frac{\Delta}{2b}$ is not changed as a result of these approximations. Then, Eq. \eqref{sol1} reads

\begin{equation}
P=P_0+\alpha_lB_lT+\frac{r}{2(r-1)}\alpha_l^2B_lT^2
\label{ml2}
\end{equation}

The ratio of quadratic and linear terms contains $\alpha_lT$ which might suggest that the quadratic term is small compared to the linear term. However, $\frac{r}{2(r-1)}$ can be large if $r=\frac{V_l}{V_s}$ is close to 1. Therefore, the quadratic term does not need to be small and should not be viewed as a next-order correction. This is consistent with the CC Eq. \eqref{e1}: $V_l\approx V_s$ gives large $\frac{dP}{dT}$. In our theory, this large $\frac{dP}{dT}$ is described by large quadratic term in Eq. \eqref{ml2}.

As the last remark in this section, we observe that equating the liquid and solid energies across the ML in Eq. \eqref{e31} (setting the small first term on the right-hand side in Eq. \eqref{e4} to zero) meant that $T$ cancelled out from all terms in Eq. \eqref{e4}, resulting in Eq. \eqref{e11}. If instead we chose not to use the equality of $E_l$ and $E_s$ and consider the right-hand side of Eq. \eqref{e3} as a new function $\Delta C_p$, $T$ would not cancel in Eq. \eqref{e11}, and $d$ would become $d=\frac{\Delta C_p}{T(V_l-V_s)}$. This would alter Eq. \eqref{e11} in two respects. First, this $d$ would include an additional and unknown {\it thermodynamic} property, $\Delta C_p$, whereas the current Eq. \eqref{e11} sets the ML in terms of thermal and elastic properties only. Second, even if we assumed $\Delta C_p$ and $V_l-V_s$ to be constant alongside constant $a$ and $b$ in Eq. \eqref{e11}, $d\propto\frac{1}{T}$ would mean that Eq. \eqref{e11} could not be integrated in a closed form (the solution would be an integral of a combination of special functions). Hence, using a physical insight regarding the equality of $E_l$ and $E_s$ significantly simplifies mathematical description and gives the parabolic form of the ML \eqref{sol1}.

\subsection{Further properties of the melting line and its relation to fundamental physical constants}

The linear term in Eq. \eqref{ml2}, $\alpha_lB_lT$, gives the following physical interpretation of the ML in terms of generic thermal and elastic effects. Lets consider a liquid just below the ML at temperature $T$. A small temperature increment $\Delta T$ gives the relative volume increase $\frac{\Delta V}{V}=\alpha_l\Delta T$, generating extra pressure $\Delta P=B_l\frac{\Delta V}{V}$. Hence,

\begin{equation}
\Delta P=\alpha_lB_l\Delta T
\label{linearslope}
\end{equation}

This combination of thermal and elastic effects increases the melting pressure at temperature $T+\Delta T$ by the amount $\Delta P$ in comparison to what this pressure would have been in the absence of these effects. This corresponds to the slope $\frac{\Delta P}{\Delta T}=\alpha_lB_l$ from Eq. \eqref{linearslope} and the linear term in Eq. \eqref{ml2}.

The same effect operates in the solid phase (our second approximation dropped this effect): negative temperature increment $\Delta T$  in the solid above the ML gives the relative volume decrease $\frac{\Delta V}{V}=\alpha_s\Delta T$, generating extra pressure $\Delta P=\alpha_sB_s\Delta T$ and corresponding to the slope $\alpha_sB_s$. The actual linear slope of the ML and $\frac{dP}{dT}$ are set by the interplay of thermal and elastic effects in both liquid and solid phases according to Eq. \eqref{e10} or \eqref{e11}.

The coefficients in front of the linear and quadratic terms of the ML parabola \eqref{ml2} are set by $\alpha_l$ and $B_l$ and, in a more general case of Eq. \eqref{e11}, by $\alpha$ and $B$ of both solids and liquids. Whereas $\alpha$ and $B$ vary in different systems, their characteristic magnitude is set by fundamental physical constants. Indeed, $B$ is set by the density of electromagnetic or bonding energy $E_b$: $B=\frac{E_b}{a_0^3}$, where $a_0$ is the interatomic separation \cite{myreview}. A shown by Weisskopf and Bernstein \cite{weisalpha}, the magnitude of $\alpha$ is set by the inverse of $E_b$: $\alpha=\frac{1}{E_b}$. %(Note that $\alpha$ is attributed to the potential asymmetry often modelled by a cubic term. This asymmetry is correlated with $E_b$ because the potential depth $E_b$ for a model asymmetric potential $U=\gamma x^2-\delta x^3$ is $E_b\propto\frac{\gamma^3}{\delta^2}$. Hence $E_b$ decreases with the potential asymmetry and, therefore, decreases with $\alpha$ - the result consistent with that of Weisskopf and Bernstein).
The scale of $E_b$ and $a_0$ is set by the Rydberg energy and the Bohr radius which, in turn, are fixed by the fundamental physical constants including the Planck constant and the electron mass $m_e$ and charge $e$. As a result, we have \cite{myreview}:

\begin{eqnarray}
\begin{split}
& B\approx\frac{E_b}{a_0^3}\propto\left(\frac{m_e^2e^5}{\hbar^4}\right)^2 \\
& \alpha\approx\frac{1}{E_b}\propto\frac{\hbar^2}{m_ee^4}
\label{fundam}
\end{split}
\end{eqnarray}

Eqs. \eqref{fundam}, together with Eq. \eqref{ml2} or \eqref{e11}, imply that (a) the parameters of the ML parabolas are governed by fundamental constants and (b) variation of ML parameters in different systems is due to system-specific proportionality coefficients in Eq. \eqref{fundam}. Hence, parabolic MLs \eqref{ml2} can not be too different in different systems and in this sense possess universality. In the next section, we will see that parameters of experimental MLs in different systems indeed fall in the characteristic range set by typical $\alpha$ and $B$.

We note that pressure typically increases $B$ and decreases $\alpha$, whereas temperature has the opposite effect. As a result, the product of $B$ and high-temperature $\alpha$ varies slower with pressure or temperature than $\alpha$ or $B$. In liquids, $\alpha_lB_l$ is in the range (3-16) $\frac{\rm MPa}{\rm K}$ as we will later see in Table 1. In solid minerals, $\alpha B$ at ambient pressure and high $T$ is in a similar range (4-7)$ \frac{\rm MPa}{\rm K}$ \cite{anderson1}. Slow variation of $\alpha B$ might be useful for developing models of temperature and pressure variations of $a$, $b$ and $d$ in Eq. \eqref{e11} and for using Eq. \eqref{ml2} where $\alpha B$ features.

Eqs. \eqref{fundam} help explain this slower variation of $\alpha B$: $B=\frac{E_b}{a^3}$ and $\alpha=\frac{1}{E_b}$ imply that $E_b$ cancels in the product $\alpha B$. As a result, this product becomes slower-varying with $P$ and $T$ and in different systems.

\subsection{Comparison to experiments}

To compare our theory to experiments and test its general applicability, we consider systems with different structure and bonding types: noble Ar and He, molecular H$_2$, network H$_2$O and metallic Fe and In. Fig. 1 shows the MLs for these systems. Consistent with experiments, Eqs. \eqref{sol} and \eqref{sol1} predict the increase of the slope of the ML with temperature. We note that low-pressure water and its ML have anomalies affected by coordination changes \cite{eisenberg}. The ML in Fig. 1 is at GPa pressures where anomalies disappear and water properties become similar to those in simple systems such as Ar \cite{nist}.

We now address the quantitative agreement between theory and experiment. In view of approximations made, this agreement is expected to be in terms of the right magnitude of ML properties rather than a close match.

\begin{figure}
\begin{center}
{\scalebox{0.36}{\includegraphics{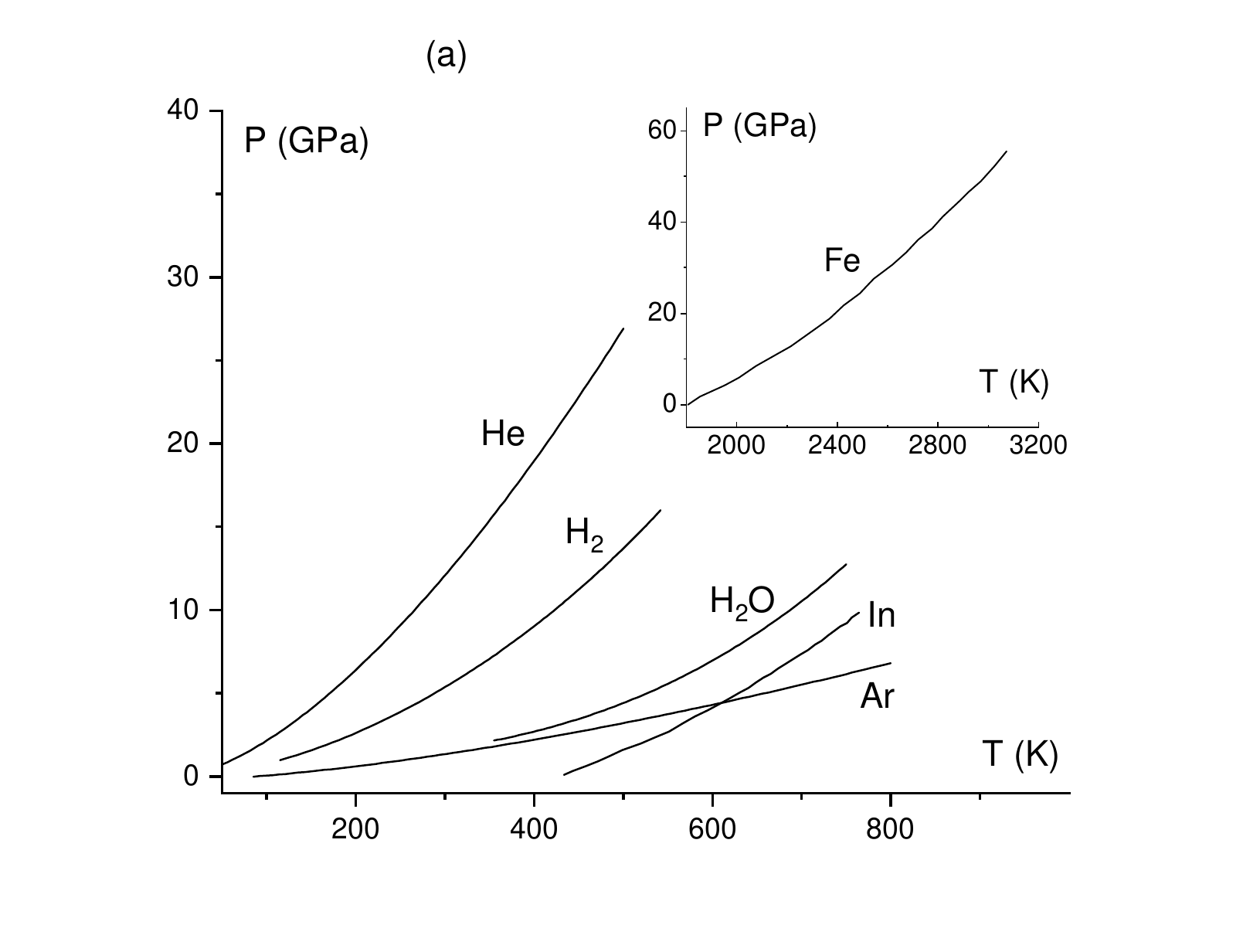}}}
{\scalebox{0.36}{\includegraphics{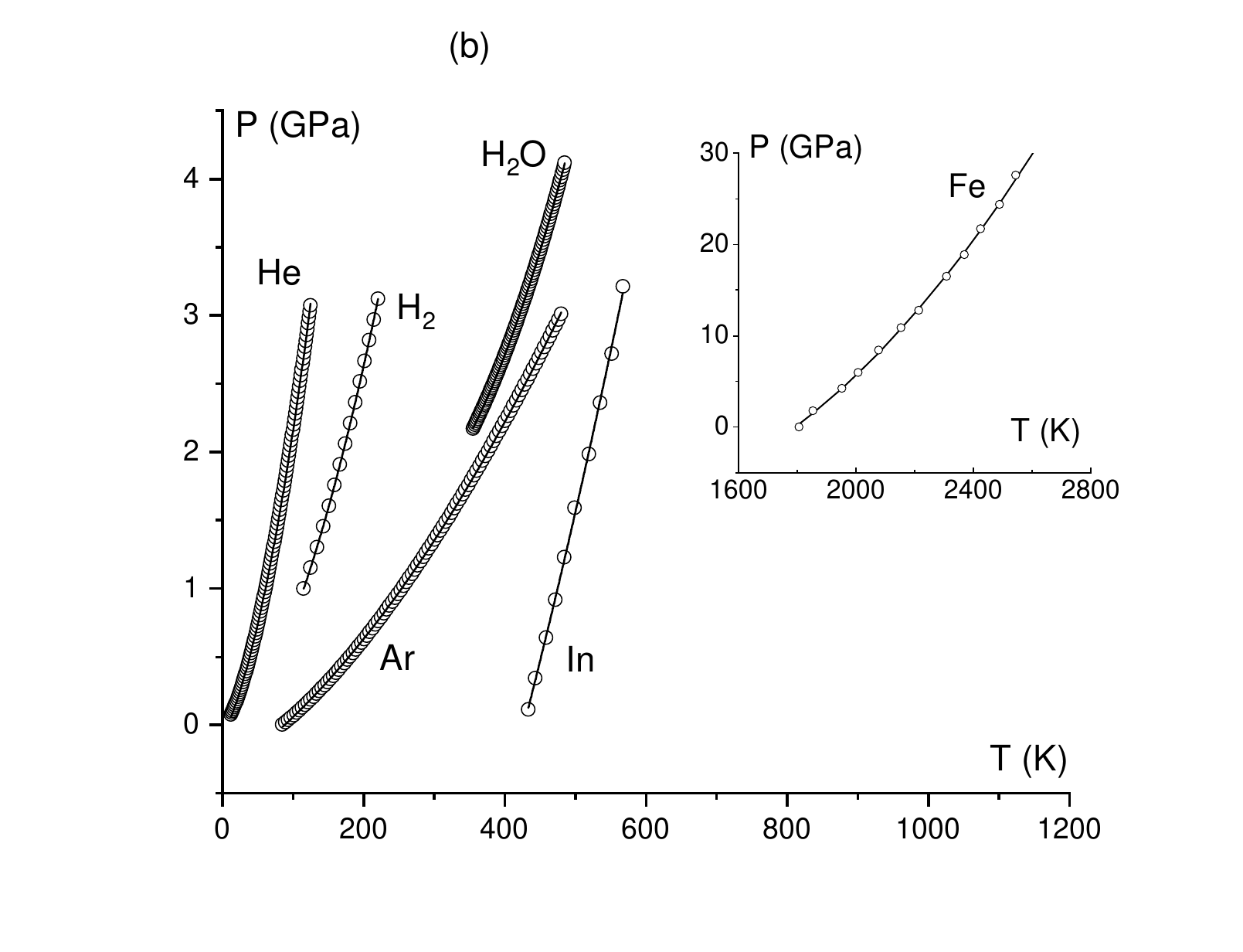}}}
\end{center}
\caption{(a): Experimental MLs of Ar, He, H$_2$ and H$_2$O \cite{datchi}. The ML for In is from an empirical EoS with experimental parameters \cite{indium}. The ML for Fe is from Ref. \cite{ironscience} shown in the inset due to a different temperature range and up to 60 GPa where static experiments exist. (b): the same MLs in the low-pressure and low-temperature range (circles). The lines show the fits to Eq. \eqref{ml2} as discussed in text.}
\label{visc}
\end{figure}

We first check whether the first linear term in Eq. \eqref{ml2} contributing $\alpha_lB_l$ to the slope of the ML gives a sensible agreement with experiments. Recall that our first approximation to derive Eq. \eqref{ml2} is related to the low pressure range where $a$, $b$ and $d$ were considered constant. Hence we calculate the experimental slope of MLs, $\frac{\Delta P}{\Delta T}$, in the low-pressure range 0-3 GPa as discussed earlier and shown in Fig. 1b. In much stiffer Fe, this range is below 100 GPa as discussed earlier and is considered in the range 0-30 GPa. Some MLs in Fig. 1b are close to linear and others are less so. In the latter case, the calculated slope is an average with a characteristic value. We compare the experimental $\frac{\Delta P}{\Delta T}$ to $2\alpha_lB_l$ in Table 1.

\begin{table}[ht]
\begin{tabular}{ l l l l l l l}
\hline
               $\alpha_lB_l$ $\left(\frac{\rm MPa}{\rm K}\right)$   & Ar & He      & H$_2$   & H$_2$O   & In   & Fe \\
 \hline
Theoretical linear slope                            & 2.5  &   6.4   & 2.8      & 3.5      & 4.1   & 16.1  \\
Experimental $\frac{\Delta P}{\Delta T}$            & 5.8  &   17.8  & 18.0     & 12.7     & 21.9  &  35.1 \\
Fits to Eq. \eqref{ml2}                             & 3.5  &  10.7   & 5.0      & 10.3     & 5.2   & 24.4  \\
\hline
\label{table}
\end{tabular}
\caption{Theoretical linear slopes of MLs, $\alpha_lB_l$, evaluated using $\alpha_l$ and $B_l$ calculated from the data in Ref. \cite{indium} for In and Ref. \cite{fealphab} for Fe in the middle of the pressure range in Fig. 1b to get characteristic values across the ML. For Ar, He, H$_2$ and H$_2$O, $\alpha_l$ and $B_l$ are calculated from the NIST data  around 1 GPa, the highest pressure at which the data is available \cite{nist}. Experimental $\frac{\Delta P}{\Delta T}$ are evaluated as average slopes of MLs in Fig. 1b. The last entry shows $\alpha_lB_l$ from the fits of experimental MLs in Fig. 1b to Eq. \eqref{ml2}.}
\end{table}

The ratios between experimental and theoretical slopes of the ML in Table 1 are about 2-3 for Ar, He, H$_2$O and Fe and 5-6 for H$_2$ and In. We note that the experimental values in the second line in Table 1 systematically exceed the theoretical linear term $\alpha_lB_l$. This is expected since, as discussed earlier, $\alpha_lB_l$ in Eq. \eqref{ml2} underestimate the linear slope in Eq. \eqref{sol}. This underestimation brings theoretical and experimental values closer.

The overall agreement indicates that the theory predicts the right order of parameters of experimental MLs and is quantitatively sensible given approximations made. In particular, the theory predicts that the slope of MLs is set by thermal and elastic properties of the system and is on the order of $\alpha_lB_l$ or about 1-10 $\frac{\rm MPa}{\rm K}$ (as opposed to, for example, $\frac{\rm kPa}{\rm K}$ or $\frac{\rm GPa}{\rm K}$).

As a second test of our theory, we fit the experimental MLs in Fig. 1b to the parabolic form of Eq. \eqref{ml2} as $P=P_0+D_1T+D_2T^2$, where $D_1$ and $D_2$ are variable. Polynomial fitting in Origin produces the fits shown in Fig. 1b with the $R$-square values very close to 1. We note that the parabolic form \eqref{ml2} can be fitted to the entire range of MLs in Fig. 1a and not only to the low pressure and temperature range in Fig. 1b where our approximation was assumed to hold.

The fitted values of $D_1$ and $D_2$ give two ways to test the theory. First, Eq. \eqref{ml2} predicts $D_1=\alpha_lB_l$. The fitted values $D_1=\alpha_lB_l$ are listed in Table 1 where we observe that the ratio of fitted and calculated values of $\alpha_lB_l$ is 1.3-1.8 for all systems except H$_2$O where this ratio is 2.9. Second, Eq. \eqref{ml2} predicts $\frac{D_2}{D_1}=\alpha_l\frac{r}{2(r-1)}$. Using $r=1.1$ as before, we find that the ratio of $\alpha_l$ obtained from fitted $\frac{D_2}{D_1}$ and $\alpha_l$ calculated using the data of Refs. \cite{nist,indium,fealphab} is 0.4-2.6 for systems in Fig. 1b.

The agreement of fitted of $\alpha_lB_l$ and $\alpha_l$ with their experimental values indicates that, similarly to the first test results, the theory gives sensible quantitative predictions of ML properties overall.

This agreement also makes a connection to the discussion in the previous section: parameters of MLs are governed by fundamental physical constants. In this section, we have seen that experimental parameters of MLs are in the characteristic range set by $\alpha$ and $B$. Indeed, the coefficient in front of the linear term in Eq. \eqref{ml2} set by $\alpha_lB_l$ is experimentally on the order of 10 $\frac{\rm MPa}{\rm K}$ in Table 1. Similarly, the experimental coefficient in front of the quadratic term is in the range set by typical $\alpha_l$ and $B_l$ as predicted. As discussed in the previous section, this characteristic range and the associated universality of ML parameters can be understood from Eq. \eqref{fundam} showing that $\alpha$ and $B$ are governed by fundamental constants.

\subsection{Relation to previous forms of the ML}
\label{otherforms}

The quadratic form of the ML in Eq. \eqref{sol1} interestingly compares to the commonly used empirical functions used to fit the ML such as the Simon-Glatzel (SG) equation of the form \cite{proctor2,chemreview}

\begin{equation}
P=D_1+D_2T^\alpha
\label{sg}
\end{equation}

\noindent where $D_1$, $D_2$ and $\alpha$ are free fitting parameters.

The fitted $\alpha$ is about 1.6 for noble Ar and He, 1.8 for H$_2$ and takes a larger value of 3 for H$_2$O \cite{datchi}. Hence this fitted $\alpha$ is often intermediate between the linear and quadratic terms in our theoretical parabolic ML in Eq. \eqref{sol1}. This is expected in view that both functions fit experimental data well. We also note that (a) differently to the empirical functions used, the theoretical ML derived here does not have free fitting parameters: as follows from Eq. \eqref{sol1} or \eqref{ml2}, the coefficients in front of linear and quadratic terms are fixed by the system thermal and elastic properties and the constant term is fixed by the location of the triple point and (b) the SG equation gives $\frac{dP}{dT}>0$ only \cite{chemreview}, whereas our theory describes anomalous melting too as discussed in Section \ref{solution}.

A similar form of the ML follows from the soft-sphere model where interactions are purely repulsive with the inverse power-law potential (IPP) $U\propto\frac{1}{r^n}$. The IPP model is thought to approximately describe interactions in simple systems such as Ar in a limited pressure range corresponding to the interatomic distances that are sufficiently shorter than the potential minimum where the attractive term is important and sufficiently longer than the short range where the IPP is too steep compared to the potential in real systems \cite{chemreview}. For the IPP, the free energy trivially scales with $VT^{\frac{3}{n}}$ \cite{landaustat}, as do other basic properties including the equation of state \cite{stishovmelt,rolandmelt}. Ref. \cite{dyremelting} explores scaling arguments applied to certain melting properties using molecular simulations.

The above scaling of the free energy and the equation of state for the IPP model results in the SG form \eqref{sg} of the ML with $\alpha=1+\frac{3}{n}$ \cite{stishovmelt,chemreview}:

\begin{equation}
P\propto T^{1+\frac{3}{n}}
\label{scaling}
\end{equation}

One might wonder how the parabolic form of the ML derived here relates to the scaling form \eqref{scaling} in the IPP model. We recall that the parabola for the ML comes from Eq. \eqref{e11} which is quite general: it follows from the CC equation \eqref{e1} using one condition only, the equality of liquid and solid $c_v$ (or energy) across the melting line as implied by theory and experiments. On the other hand, the IPP model is a peculiar distinctive system constrained by several rigid scaling conditions. Two relevant constraints are: (a) the entropy change across the ML is constant, $\Delta S$=const (compare this with the condition $T\Delta S$=const used to derive boiling and sublimation lines, see Introduction), and (b) the volumes of both phases scale as $V\propto T^{-\frac{3}{n}}$ along the ML \cite{stishovmelt}. These constraints change the differential equation for the ML. In particular, the right-hand side of Eq. \eqref{e2} becomes zero and derivatives $\frac{dV}{dT}$ in the second term simplify and become $\frac{dV}{dT}\propto -\frac{3}{n}T^{-\frac{3}{n}-1}$. The differential equation \eqref{e2} for the IPP model then becomes

\begin{equation}
\frac{d^2P}{dT^2}=\frac{3}{n}\frac{1}{T}\frac{dP}{dT}
\label{diffscale}
\end{equation}

\noindent and is different from Eq. \eqref{e11} governing the ML.

Eq. \eqref{scaling} for the ML follows from Eq. \eqref{diffscale}. The same Eq. \eqref{scaling} follows if the two constraints from the IPP model, $\Delta S$=const and $V\propto T^{-\frac{3}{n}}$, are used in the CC equation \eqref{e101}.

It is therefore an expected result that the derived parabolic ML differs from the scaling form \eqref{scaling}: by imposing scaling constraints, the IPP model changes the differential equation for the ML and its solution.

In passing, we note that $\alpha=1+\frac{3}{n}$ in Eq. \eqref{scaling} is $\alpha=1.25$ if a typical value $n=12$ is used \cite{hooverscaling} and $\alpha=1.17$ if $n$ is taken as $n=18$ ($n=18$ is considered as an effective exponent of the IPP approximating the Lennard-Jones (LJ) potential \cite{eff18}). These $\alpha$ are significantly smaller than $\alpha=1.6$ experimentally seen in Ar and He \cite{datchi} where the LJ potential is considered to operate and where the IPP is assumed to approximately apply at high pressure. To be consistent with the experimental $\alpha=1.6$, $n$ in Eq. \eqref{scaling} should be $n=5$, which is significantly smaller than what's expected for the IPP in real systems. The origin of this discrepancy is unclear. This is especially so in view that the scaling property $V\propto T^{-\frac{3}{n}}$ of the IPP model applies to the ML of Ar well with $n$ close to expected $n=12$ \cite{stishovdensity}. A potentially useful observation is that the pressure range in these experiments \cite{stishovdensity,datchi} was different. The earlier experiments yielding $n=12$ from density scaling in Ar were up to about 1 GPa \cite{stishovdensity}, whereas the $P(T)$ ML was fitted to Eq. \eqref{scaling} with $\alpha=1.6$ ($n=5$) up to 50 GPa, or over 10,000 times the critical pressure \cite{datchi}. In He, the ML was fitted to Eq. \eqref{scaling} with $\alpha=1.6$ up to 24 GPa, or over 100,000 times the critical pressure (the ML could not be fitted at higher pressure up to 42 GPa) \cite{datchi}. This difference of pressure ranges could affect the fitted exponents, which would be qualitatively consistent with the variability of the effective $n$ with the state point for the LJ potential \cite{pederseneff}. This then raises a more general question of the extent to which earlier models including the IPP model applies to real physical systems and MLs in particular.

%Pc 4.86 MPa in Ar and 0.23 MPa in He

\section{Final remarks and summary}

We make two final remarks. First, recall that the properties of the sublimation and boiling lines are derived by assuming that the latent heat $q$ in the CC equation \eqref{e1} is constant \cite{landaustat,kittel}. %Using this assumption for the ML would give $\frac{dP}{dT}=\frac{q}{(V_l-V_s)}\frac{1}{T}\propto\frac{1}{T}$, where we neglected next-order thermal expansion effects in the term $V_l-V_s$. This gives the slope of the ML decreasing with $T$, in contrast to experimental MLs in Fig. 1. Hence {\it not}
Not assuming constant $q$ and considering a more general case was important in our theory, in Eqs. \eqref{e101}, \eqref{e2} and all equations that followed. This suggests that our theory can be applied to those parts of the sublimation and boiling lines where $q$ is not constant and where the textbook solution of the CC equation, $P(T)\propto e^{-\frac{q}{T}}$ \eqref{plines}, deviates from experiments \cite{reidexp}. This would also enable to discuss these two lines in more general terms and improve their understanding.

Second, the derived parabolic line for the melting line can be used to draw the Frenkel line in the supercritical state \cite{flreview}. Indeed, the Frenkel line starts just below the critical point and runs approximately parallel to the melting line in the double-logarithmic plot because both lines correspond to the qualitative change of particle dynamics \cite{Yang2015}.

In summary, we proposed a general theory of the ML for the first time. This theory describes the ML in terms of thermal and elastic properties of liquid and solid phases. We showed that the approximate solutions of the theory are in agreement with experimental MLs. The agreement can be refined by fully solving Eq. \eqref{e11} with variable $a$, $b$ and $d$ in future work. We also showed that the parameters of the ML parabola are governed by the fundamental constants. For this reason, MLs can not be too different in different systems and in this sense possess universality.

I am grateful to V. Brazhkin, J. Dyre, J. Proctor, P. Tello and G. de With for discussions and EPSRC for support.

%\bibliography{refs}

\bibliographystyle{apsrev4-1}

\end{document}